\documentclass[AMA,Times1COL]{WileyNJDv5} 

\articletype{Short Communication}%

\received{Date Month Year}
\revised{Date Month Year}
\accepted{Date Month Year}
\journal{Software: Practice and Experience}
\volume{00}
\copyyear{2024}
\startpage{1}

\algnewcommand{\IfThenElse}[3]{
  \State \algorithmicif\ #1\ \algorithmicthen\ #2\ \algorithmicelse\ #3}
  
\algnewcommand{\IfThen}[2]{
  \State \algorithmicif\ #1\ \algorithmicthen\ #2}

\algnewcommand{\OneLineFor}[2]{
  \State \algorithmicfor\ #1\ \algorithmicdo\ #2}

\begin{document}

\title{Algorithms for Generating Small Random Samples}

\author[1]{Vincent A Cicirello}

\authormark{Vincent A Cicirello}
\titlemark{Algorithms for Generating Small Random Samples}

\address[1]{\orgdiv{Computer Science}, \orgname{Stockton University}, \orgaddress{\state{New Jersey}, \country{USA}}}

\corres{Vincent A Cicirello, Computer Science, School of Business, Stockton University, 101 Vera King Farris Dr, Galloway, NJ 08205 \email{vincent.cicirello@stockton.edu}}

\presentaddress{Vincent A Cicirello, Computer Science, School of Business, Stockton University, 101 Vera King Farris Dr, Galloway, NJ 08205 \email{vincent.cicirello@stockton.edu}}

\abstract[Abstract]{We present algorithms for generating small
random samples without replacement. We consider two cases. We 
present an algorithm for sampling a pair of distinct integers, 
and an algorithm for sampling a triple of distinct integers. The 
worst-case runtime of both algorithms is constant, while the 
worst-case runtimes of common algorithms for the general case of 
sampling $k$ elements from a set of $n$ increase with $n$. Java 
implementations of both algorithms are included in the open 
source library $\rho\mu$.}


\maketitle

\renewcommand\thefootnote{}
\footnotetext{\textbf{Abbreviations:} JMH, Java microbenchmark harness; 
JVM, Java virtual machine; PRNG, pseudorandom number generator.}

\renewcommand\thefootnote{\fnsymbol{footnote}}
\setcounter{footnote}{1}

\section{Introduction}

Efficiently generating random samples of $k$ elements from a set of $n$
elements, without replacement, is important to a variety of applications.
There are algorithms for handling this general case, such as the reservoir
sampling family of algorithms~\cite{Vitter1985,Li1994}, pool 
sampling~\cite{Goodman1977}, insertion sampling~\cite{cicirello2022applsci}, 
among others~\cite{Ernvall1982,Knuth2}. The runtime of pool sampling and
some forms of reservoir sampling is $O(n)$. 
Other forms of reservoir sampling~\cite{Li1994} improve the 
runtime to $O(k(1 + \ln(n/k)))$, but this still increases with $n$.
Insertion sampling was designed with small samples in mind, with a 
runtime of $O(k^2)$, requiring $k$ random numbers to generate a sample of size 
$k$. It is an efficient choice when you know $k$ will be small, although for 
large $k$ it is much slower than alternatives.
Others such as Ting make other improvements to the 
classical random sampling algorithms~\cite{Ting2021}. There are also
several parallel algorithms for random sampling without replacement~\cite{Sanders2018},
weighted random sampling~\cite{HubschleSchneider2022}, and for random sampling 
(both weighted and unweighted) from data streams~\cite{Tangwongsan2019,Karras2022,HbschleSchneider2020}.

Although insertion sampling handles small $k$ nicely, we can do much 
better with algorithms designed for the specific value of $k$. 
For example, the special case of $k=1$ has been widely 
studied, such as Lemire's algorithm for random integers in an interval 
that is significantly faster than the alternatives provided within common
programming languages by avoiding nearly all division operations~\cite{Lemire2019}, 
or Goualard's approach to sampling random floating-point numbers from an 
interval~\cite{Goualard2022}. It is common for applications to require 
repeatedly generating random integers from an interval, such as algorithms
for shuffling arrays, as well as sampling algorithms for the general case.
Brackett-Rozinsky and Lemire propose an algorithm for sampling multiple 
independent bounded random integers from a single random word, and demonstrate how the 
technique can speed up algorithms for shuffling~\cite{BrackettRozinsky2024}.

In this article, we present algorithms for the special cases of $k=2$ and $k=3$. 
The runtime of both algorithms is $O(1)$, with the algorithm for generating a random pair
of distinct integers requiring two random numbers, and the algorithm for generating
a random triple of distinct integers requiring three random numbers. Our original 
motivation for efficient algorithms for randomly sampling pairs and triples of 
distinct integers is in the implementation of various mutation and crossover 
operators for evolutionary algorithms for the space of 
permutations~\cite{cicirello2023ecta}, where such evolutionary operators require 
generating random combinations of indexes. In such an application, one requires
random samples from the integer interval $[0,n)$. Thus, the algorithms that we 
present are specified as sampling from that interval. However, without loss
of generality, both algorithms are applicable for the broader case of sampling pairs 
and triples of elements from a set of $n$ elements, provided you define a mapping of
the $n$ elements to the integers in $[0,n)$.
We utilize Lemire's algorithm for bounded random integers~\cite{Lemire2019}
within our implementation of the algorithms of this paper, as well as in our
implementations of the general sampling algorithms used in the experiments.
The Brackett-Rozinsky and Lemire approach to multiple bounded random 
integers~\cite{BrackettRozinsky2024} can potentially enable further optimizations
as well, but we have not explored doing so.

The two algorithms for sampling pairs and triples of integers are presented in
Section~\ref{sec:algs}. 
We also discuss in that section how one can derive a special
purpose sampling algorithm for any fixed $k$ in the form of a sampling network,
although for larger $k$ this is likely impractical, and demonstrate with an
algorithm for the case of $k=4$.
We provide Java implementations in the open source library $\rho\mu$, which 
provides a variety of randomization-related utilities~\cite{cicirello2022joss}. 
We empirically compare the runtime performance of the algorithms to existing 
random sampling algorithms using the $\rho\mu$ library, describing our experimental 
methodology in Section~\ref{sec:methodology} and the results in Section~\ref{sec:results}. 
We wrap up in Section~\ref{sec:conclusion}.

\section{Algorithms}\label{sec:algs}

We now present the algorithms. In our pseudocode, the function $\mathrm{Rand}(n)$
returns a random integer uniformly distributed in the half-open interval $[0,n)$.

Algorithm~\ref{alg:pair}, $\mathrm{RandomPair}(n)$, generates a random pair $(i,j)$ of
distinct integers from the interval $[0,n)$, uniformly distributed over the space
of all $2 {n \choose 2}$ distinct pairs. Line 1 generates the 
first of the two integers, $i$, uniformly from the interval $[0,n)$. Then, a random 
integer $j$ uniformly distributed over the interval $[0,n-1)$ is generated (line 2);
and if $j$ is the same as $i$, we reset $j$ to $n-1$ (line 3), since $n-1$ was 
excluded via the half-open interval $[0,n-1)$.

\begin{algorithm}[t]
\caption{$\mathrm{RandomPair}(n)$}
\begin{algorithmic}[1]
\State $i \leftarrow \mathrm{Rand}(n)$
\State $j \leftarrow \mathrm{Rand}(n-1)$
\IfThen{$j = i$}{$j \leftarrow n - 1$}
\State \textbf{return} $(i, j)$
\end{algorithmic}
\label{alg:pair}
\end{algorithm}

Algorithm~\ref{alg:triple}, $\mathrm{RandomTriple}(n)$, generates a random triple
$(i, j, k)$ of distinct integers from the interval $[0,n)$, uniformly distributed 
over the space of all $6 {n \choose 3}$ distinct triples. It initially 
generates $i$, $j$, and $k$ uniformly at random from the half-open intervals 
$[0,n)$, $[0,n-1)$, and $[0,n-2)$, respectively. The remainder of the algorithm maps 
duplicates to the values excluded by the half-open intervals. 
Line 4 ensures that $j$ and $k$ are different. Lines 5--6 in turn
adjusts if necessary so that neither are the same as $i$.

\begin{algorithm}[t]
\caption{$\mathrm{RandomTriple}(n)$}
\begin{algorithmic}[1]
\State $i \leftarrow \mathrm{Rand}(n)$
\State $j \leftarrow \mathrm{Rand}(n-1)$
\State $k \leftarrow \mathrm{Rand}(n-2)$
\IfThen{$k = j$}{$k \leftarrow n - 2$}
\IfThen{$j = i$}{$j \leftarrow n - 1$}
\IfThen{$k = i$}{$k \leftarrow n - 1$}
\State \textbf{return} $(i, j, k)$
\end{algorithmic}
\label{alg:triple}
\end{algorithm}

The worst-case runtime of both algorithms is $O(1)$. 
Algorithm~\ref{alg:pair} and Algorithm~\ref{alg:triple} 
generate 2 and 3 random integers, respectively; and
both have constant numbers of comparisons and assignments 
to adjust for duplicates.
 
The bounds of the random numbers utilized by Algorithm~\ref{alg:pair} 
and Algorithm~\ref{alg:triple} are similar to those of the first two 
and three steps of a Fisher-Yates shuffle~\cite{Durstenfeld1964} as well
as the first two to three iterations of insertion sampling (see later in
Section~\ref{sec:other}). Additionally,
each if-statement that resolves duplicates is reminiscent of
the ``compare-exchange'' operation of sorting 
networks~\cite{Batcher1968}. But, instead of a ``compare-exchange,'' 
it is a ``compare-change.'' This suggests the potential to
derive a {\em sampling network} from the approach. Such a
sampling network may have similar benefits as sorting networks,
such as lending themselves well to hardware implementation
and parallelism. For example, the execution order of lines 5 
and 6 of Algorithm~\ref{alg:triple} does not matter. It is
straightforward to derive a corresponding sampling network
for any fixed $k$, although the benefit in software diminishes
since the obvious approach leads to a number of compare-changes
that increases quadratically in $k$. For example, 
Algorithm~\ref{alg:quad} shows how to extend the approach to $k=4$.
A hypothetical circuit implementation could align the compare-changes
in layers of independent operations, such as lines 6--7 in parallel
and lines 8--10 in parallel.

\begin{algorithm}[t]
\caption{$\mathrm{RandomFourTuple}(n)$}
\begin{algorithmic}[1]
\State $h \leftarrow \mathrm{Rand}(n)$
\State $i \leftarrow \mathrm{Rand}(n-1)$
\State $j \leftarrow \mathrm{Rand}(n-2)$
\State $k \leftarrow \mathrm{Rand}(n-3)$
\IfThen{$k = j$}{$k \leftarrow n - 3$}
\IfThen{$j = i$}{$j \leftarrow n - 2$}
\IfThen{$k = i$}{$k \leftarrow n - 2$}
\IfThen{$i = h$}{$i \leftarrow n - 1$}
\IfThen{$j = h$}{$j \leftarrow n - 1$}
\IfThen{$k = h$}{$k \leftarrow n - 1$}
\State \textbf{return} $(h, i, j, k)$
\end{algorithmic}
\label{alg:quad}
\end{algorithm}

\section{Experimental Methodology}\label{sec:methodology}

\subsection{General Sampling Algorithms}\label{sec:other}

We experimentally validate the approach relative to existing algorithms
for the general case of sampling $k$ distinct elements from a set of $n$
to explore the benefit of special purpose algorithms when $k$ is known
and small.

First, we consider two forms of reservoir sampling~\cite{Vitter1985,Li1994}, 
which refers to a family of sampling algorithms. Reservoir algorithms were
originally introduced decades ago for the problem of randomly sampling $k$
records from an unknown number of records $n$ via a single pass over the 
records. But they are also easily adapted to the simpler problem of sampling
integers with known $n$. The first reservoir sampling algorithm that we 
consider is the simplest, and referred to by Vitter as algorithm 
R~\cite{Vitter1985}. We provide pseudocode in Algorithm~\ref{alg:R}.
It begins by initializing the sample with the first $k$ elements (lines 1--2).
It then iterates over the remaining $n-k$ elements in a very similar way
to an array shuffling procedure, choosing a random index $j$ for the
next element (lines 3--4). If that index is in the bounds of the sample,
that element replaces the one at that index in the sample (line 5). The
runtime of this reservoir sampling algorithm is $O(n)$, and it requires
$O(n-k)$ random numbers, which is desirable when $k$ is large, but there
are better algorithms for small $k$. In the same article, Vitter described
a series of improvements leading to his algorithms X, Y, and Z, the last
of which having an expected runtime of $O(k(1 + \ln(n/k)))$.

\begin{algorithm}[t]
\caption{$\mathrm{ReservoirSamplingR}(n, k)$}
\begin{algorithmic}[1]
\State $\mathrm{sample} \leftarrow \text{a new array of length}\ k$
\OneLineFor{$i = 0$ \textbf{to} $k-1$}{$\mathrm{sample}[i] \leftarrow i$}
\For{$i = k$ \textbf{to} $n-1$}
\State $j \leftarrow \mathrm{Rand}(i+1)$
\IfThen{$j<k$}{$\mathrm{sample}[j] \leftarrow i$}
\EndFor
\State \textbf{return} $\mathrm{sample}$
\end{algorithmic}
\label{alg:R}
\end{algorithm}

The other reservoir sampling algorithm that we consider is Li's algorithm
L~\cite{Li1994}, which is simpler and an improvement over Vitter's algorithm 
Z~\cite{Vitter1985}. We provide pseudocode for the L form of reservoir
sampling in Algorithm~\ref{alg:L}. In the pseudocode, $U()$ returns a
random floating-point value in $[0.0,1.0)$. It begins like the previous reservoir
algorithm by initializing the sample with the first $k$ elements (lines 1--2).
Instead of explicitly iterating over all $n-k$ remaining elements, it
skips groups of elements. See Li for derivation of the skip lengths
and the details of the runtime analysis~\cite{Li1994}. The expected runtime 
still grows with $n$, but much better than linearly, $O(k(1 + \ln(n/k)))$.
As $k$ approaches $n$, this converges to $O(k)$. Thus, this is an especially
good choice for larger $k$. There are $k$ steps to initialize the sample, and 
the expected number of iterations of the while loop is $O(k \ln(n/k))$. It 
should be noted that the iterations are more costly with each iteration generating 
two uniform random floating-point values and a random bounded integer, along with 
an exponentiation and three logs.

\begin{algorithm}[t]
\caption{$\mathrm{ReservoirSamplingL}(n, k)$}
\begin{algorithmic}[1]
\State $\mathrm{sample} \leftarrow \text{a new array of length}\ k$
\OneLineFor{$i = 0$ \textbf{to} $k-1$}{$\mathrm{sample}[i] \leftarrow i$}
\State $w \leftarrow \exp\left(\frac{\ln(U())}{k}\right)$
\State $i \leftarrow k + \left\lfloor\frac{\ln(U())}{\ln(1-w)}\right\rfloor$
\While{i < n}
  \State $\mathrm{sample}[\mathrm{Rand}(k)] \leftarrow i$
  \State $w \leftarrow w \cdot \exp\left(\frac{\ln(U())}{k}\right)$
  \State $i \leftarrow i + 1 + \left\lfloor\frac{\ln(U())}{\ln(1-w)}\right\rfloor$
\EndWhile
\State \textbf{return} $\mathrm{sample}$
\end{algorithmic}
\label{alg:L}
\end{algorithm}

Next consider pool sampling~\cite{Goodman1977}, which has additional names in the
literature, including random draw sampling~\cite{Bellhouse1991}. Its authors originally
referred to it generically as ``SELECT''~\cite{Goodman1977}, but there are others with 
that same name. Algorithm~\ref{alg:pool} shows it in pseudocode form. It maintains a
pool of unused elements (initialized in lines 2--4 and updated in lines 8--9), and
draws elements randomly from the pool (lines 6--7) until the sample is complete. Pool 
sampling requires $k$ random numbers to generate a sample of size $k$, which is good 
when $k$ is small, however its runtime grows linearly in $n$; and it also requires 
$O(n)$ extra storage. Ernvall and Nevalainen~\cite{Ernvall1982} use a hash table to 
reduce the extra storage to $O(k)$ and the runtime to $O(k^2)$.

\begin{algorithm}[t]
\caption{$\mathrm{PoolSampling}(n, k)$}
\begin{algorithmic}[1]
\State $\mathrm{sample} \leftarrow \text{a new array of length}\ k$
\State $\mathrm{pool} \leftarrow \text{a new array of length}\ n$
\OneLineFor{$i = 0$ \textbf{to} $n-1$}{$\mathrm{pool}[i] \leftarrow i$}
\State $\mathrm{remaining} \leftarrow n$
\For{$i = 0$ \textbf{to} $k-1$}
  \State $j \leftarrow \mathrm{Rand}(\mathrm{remaining})$
  \State $\mathrm{sample}[i] \leftarrow \mathrm{pool}[j]$
  \State $\mathrm{remaining} \leftarrow \mathrm{remaining} - 1$
  \State $\mathrm{pool}[j] \leftarrow \mathrm{pool}[\mathrm{remaining}]$
\EndFor
\State \textbf{return} $\mathrm{sample}$
\end{algorithmic}
\label{alg:pool}
\end{algorithm}

Like pool sampling, the insertion sampling~\cite{cicirello2022applsci} 
algorithm requires only $k$ random numbers. But unlike pool sampling, 
its storage requirement is constant. Its runtime is $O(k^2)$. It is
not widely studied as it was introduced within the context of a specific
application, one where $k$ is unknown ahead of time but likely small.
Algorithm~\ref{alg:insertion} provides the details. It maintains the 
currently sampled elements in sorted order (lines 5, 7--10). It iteratively 
generates a sequence of random integers 
$v \in [\mathrm{Rand}(n), \mathrm{Rand}(n-1), \ldots, \mathrm{Rand}(n-k+1)]$
(line 3). Each $v$ is treated as an index into an ordered (implicit) list of
the unused elements. Line 6 skips over used elements as $v$ is inserted
into the sample in its sorted position (line 10).

\begin{algorithm}[t]
\caption{$\mathrm{InsertionSampling}(n, k)$}
\begin{algorithmic}[1]
\State $\mathrm{sample} \leftarrow \text{a new array of length}\ k$
\For{$i = 0$ \textbf{to} $k-1$}
  \State $v \leftarrow \mathrm{Rand}(n - i)$
  \State $j \leftarrow k - i$
  \While{$j < k$ \textbf{and} $v \geq \mathrm{sample}[j]$}
    \State $v \leftarrow v + 1$
	\State $\mathrm{sample}[j-1] \leftarrow \mathrm{sample}[j]$
	\State $j \leftarrow j + 1$
  \EndWhile
  \State $\mathrm{sample}[j-1] \leftarrow v$
\EndFor
\State \textbf{return} $\mathrm{sample}$
\end{algorithmic}
\label{alg:insertion}
\end{algorithm}

The proposed algorithms in Section~\ref{sec:algs} 
(Algorithms~\ref{alg:pair},~\ref{alg:triple}, and~\ref{alg:quad})
produce samples that are not only uniformly random over combinations,
but also uniformly random over permutations. This is also true of
pool sampling (Algorithm~\ref{alg:pool}). However, it is not true
of the other reference algorithms. Insertion sampling 
(Algorithm~\ref{alg:insertion}) generates an ordered sample. For 
uniform ordering, shuffle the sample before the return on line 12. 
Line 2 of both versions of reservoir sampling
(Algorithms~\ref{alg:R} and~\ref{alg:L}) constrains the possible
positions of the first $k$ elements. Shuffling the initial sample 
of line 2 achieves uniform random ordering. In all three cases, 
the adjustment introduces an $O(k)$ step with $O(k)$ additional
random numbers. In the experiments, we use these reference algorithms 
as specified in the pseudocode and do not introduce such shuffling steps.

\subsection{Methodology}

We implemented the new algorithms from Section~\ref{sec:algs}, and three of the
other sampling algorithms from Section~\ref{sec:other} in the open source library 
$\rho\mu$. The experiments use $\rho\mu$ 4.2.0. The code for 
reservoir (L) sampling is available in the GitHub repository that contains
the code to reproduce the experiments, and the raw data from my runs. See 
Table~\ref{tab:rhomu} for URLs to the library source code, project website, and in 
the Maven Central repository, as well as for the source code and raw data from 
the experiments.

\begin{table}[t]
\caption{Important URLs for the $\rho\mu$ open source library and experiments}\label{tab:rhomu}
\centering
\begin{tabular*}{\textwidth}{@{}ll@{\extracolsep\fill}}\toprule
\multicolumn{2}{@{}l}{$\rho\mu$ library} \\\midrule
Source & \url{https://github.com/cicirello/rho-mu} \\
Website	& \url{https://rho-mu.cicirello.org/} \\
Maven & \url{https://central.sonatype.com/artifact/org.cicirello/rho-mu/} \\\midrule[\heavyrulewidth]
\multicolumn{2}{@{}l}{Experiments} \\\midrule
Source/data & \url{https://github.com/cicirello/small-sample-experiments} \\\bottomrule
\end{tabular*}
\end{table}

Table~\ref{tab:methods} lists the methods of $\rho\mu$'s \lstinline|EnhancedRandomGenerator| 
class that are used in the experiments. The \lstinline|EnhancedRandomGenerator| class wraps 
any class that implements Java 17's \lstinline|RandomGenerator| interface adding a variety 
of functionality to it, or in some cases replacing implementations with more efficient
algorithms. There are multiple \lstinline|nextIntPair| and \lstinline|nextIntTriple| methods, 
which differ in terms of return type (e.g., returning an array of type \lstinline|int| 
versus returning a library specific record \lstinline|IndexPair| and \lstinline|IndexTriple|).

\begin{table}[t]
\caption{Methods of $\rho\mu$'s \lstinline|EnhancedRandomGenerator| class used in
the experiments}\label{tab:methods}
\centering
\begin{tabular*}{\textwidth}{@{}ll@{\extracolsep\fill}}\toprule
Algorithm & Method of \lstinline|EnhancedRandomGenerator| class \\ \midrule
\lstinline|RandomPair| (Algorithm~\ref{alg:pair}) & \lstinline|nextIntPair| \\
\lstinline|RandomTriple| (Algorithm~\ref{alg:triple}) & \lstinline|nextIntTriple| \\
insertion sampling~\cite{cicirello2022applsci} & \lstinline|sampleInsertion| \\
pool sampling~\cite{Goodman1977} & \lstinline|samplePool| \\
reservoir (R) sampling~\cite{Vitter1985} & \lstinline|sampleReservoir| \\ 
reservoir (L) sampling~\cite{Li1994} & \lstinline|ReservoirL| class in experiment repository \\
\bottomrule
\end{tabular*}
\end{table}

We use OpenJDK 64-Bit Server VM version 17.0.2 in the experiments on a Windows 10 PC
with an AMD A10-5700, 3.4 GHz processor and 8GB memory. For the pseudorandom number
generator (PRNG), we use Java's \lstinline|SplittableRandom| class, which implements the
SplitMix~\cite{Steele2014} algorithm, which is a faster optimized version of the 
DotMix~\cite{Leiserson2012} algorithm, and which passes the 
DieHarder~\cite{brown2013dieharder} tests.

The $\rho\mu$ library's \lstinline|EnhancedRandomGenerator| class replaces Java's
method for generating random integers subject to a bound with an implementation of
a much faster algorithm~\cite{Lemire2019}. Our prior experiments~\cite{cicirello2022joss} 
show that Lemire's approach uses less than half the CPU time as Java's
\lstinline|RandomGenerator.nextInt(bound)| method. So although we are using Java's
\lstinline|SplittableRandom| class as the PRNG, by wrapping it in an instance of
$\rho\mu$'s \lstinline|EnhancedRandomGenerator|, we are significantly speeding the runtime
of all of the algorithms in our experiments.

We use the Java Microbenchmark Harness (JMH)~\cite{JMH137} to implement our experiments.
JMH is developed by the same team that develops the OpenJDK, and is
the preferred harness for benchmarking Java code. Since the Java Virtual Machine (JVM) is 
adaptive, utilizing just-in-time 
compilation of hot-spots identified at runtime to boost performance, it is important
to warm up the JVM to achieve a steady-state before benchmarking. The JMH handles this
warmup phase, and other Java benchmarking related challenges.
For each experiment condition (e.g., algorithm and value of $n$) we use five 10-second warmup 
iterations to ensure that the Java JVM is properly warmed up, and we likewise use
five 10-second iterations for measurement. 
JMH executes the benchmark method as many times as each 10-second 
iteration allows, which leads to hundreds of millions of method invocations for each
experiment condition. We use JMH's built-in blackhole to consume the generated samples
to prevent dead-code elimination.
We measure and report average time per
operation in nanoseconds, along with 99.9\% confidence intervals 
as calculated by JMH. The times reported by JMH are not CPU times.
Although not indicated in the JMH documentation, examination of the JMH
source code~\cite{JMH137} reveals that it calculates elapsed time using the
JVM's high-resolution time source via the \lstinline|System.nanoTime()| method,
which returns nanoseconds since an arbitrary origin. We round the averages to the 
nearest tenth of a nanosecond since the JVM's time source is of nanosecond precision
(JMH's reported averages are to thousandths of a nanosecond, a false level of precision). 
A consequence of elapsed time is that benchmarks can be affected by the state of
the system (e.g., competition for system resources). To minimize such impact, we shut 
down all applications under our control, and disabled various non-critical background
processes (e.g., auto-updaters, etc) to limit background processes as much as feasible.

\section{Results}\label{sec:results}

\subsection{Sampling Pairs of Distinct Integers}

This first set of results in Table~\ref{tab:pairs} explores the difference in 
runtime performance of the various approaches to sampling random pairs of 
distinct integers from the half open interval $[0, n)$. It is clear from the
results that the \lstinline|RandomPair| (Algorithm~\ref{alg:pair}) of this 
paper dominates all of the others. Its average runtime is a low constant, and
does not vary as $n$ increases, at least not to any significant degree. If it
appears that average time is possibly decreasing as $n$ increases for this 
algorithm, it may be, although it is clearly negligible. This is because as 
$n$ increases, the probability of the conditional assignment on line 3 of 
Algorithm~\ref{alg:pair} decreases. The insertion sampling~\cite{cicirello2022applsci} 
average runtime is likewise approximately constant, though a higher constant. Note 
that insertion sampling's theoretical runtime is $O(k^2)$, which is essentially 
constant for any fixed $k$. Pool sampling~\cite{Goodman1977} and reservoir (R) 
sampling~\cite{Vitter1985} both have an average runtime that increases linearly 
in $n$. Reservoir (R) sampling's average runtime is much higher than pool sampling, 
however, because reservoir (R) sampling also generates $(n-k)$ random integers to 
sample $k$ integers from a set of $n$ integers, while pool sampling only needs 
$k$ random integers to accomplish this. Thus, reservoir (R) is an especially poor 
choice for low $k$ relative to $n$. 
Reservoir (L) sampling~\cite{Li1994} is the 
slowest for low $n$, but it overtakes reservoir (R) as $n$ increases, and should
surpass the performance of pool sampling if we further increase $n$. Its
runtime increases with $n$, but at a slower rate $O(k(1 + \ln(n/k)))$ than 
reservoir (R) and pool sampling.

\begin{table}[t]
\caption{Sampling pairs of distinct integers: average time (ns)
with 99.9\% confidence intervals}\label{tab:pairs}
\centering
\begin{tabular*}{\textwidth}{@{\extracolsep\fill}rrrrrr@{}}\toprule
$n$	& \lstinline|RandomPair| (Algorithm~\ref{alg:pair}) & insertion~\cite{cicirello2022applsci} & pool~\cite{Goodman1977} & reservoir (R)~\cite{Vitter1985} & reservoir (L)~\cite{Li1994} \\ \midrule
$n=16$   & $16.0 \pm 0.5$    & $24.1 \pm {<}0.1$ & $54.9 \pm 0.3$    & $178.2 \pm 1.0$   & $465.8 \pm 14.5$\\
$n=64$   & $15.6 \pm 0.1$    & $24.1 \pm 0.1$    & $89.3 \pm 1.9$    & $633.4 \pm 9.7$   & $721.3 \pm 19.5$\\
$n=256$  & $15.7 \pm 0.6$    & $24.3 \pm {<}0.1$ & $298.8 \pm 0.8$   & $2062.5 \pm 9.7$  & $981.3 \pm 8.4$\\
$n=1024$ & $15.7 \pm {<}0.1$ & $23.6 \pm 0.1$    & $1209.9 \pm 16.2$ & $7918.2 \pm 12.9$ & $1242.7 \pm 1.4$\\
\bottomrule
\end{tabular*}
\end{table}

The results discussed above used the version of the \lstinline|EnhancedRandomGenerator.nextIntPair|
method that allocates and returns an array for the pair of integers, which is also
what the various sampling algorithm implementations do. The $\rho\mu$ library also provides
a version of the \lstinline|EnhancedRandomGenerator.nextIntPair| that returns a Java record.
A Java record is immutable and potentially enables the JVM to optimize in ways that it otherwise
cannot. To explore this, we ran an additional experiment comparing three cases: (a) allocating
and returning a new array for the integer pair, (b) using a preallocated array passed as a
parameter, and (c) allocating and returning a record instead of an array. In this experiment,
rather than using JMH's blackhole to consume the generated pairs directly (as we did in the 
first set of results), we simulate an operation on the pair of integers to give the JVM the
opportunity to exploit the immutability property of Java records. Specifically, we do a simple
sum of the integers in the pairs, and return that to JMH's blackhole.

Table~\ref{tab:pairs-ds} shows the results of this second experiment. The fastest of the 
three options uses a Java record type for the pair of integers. It was even faster than 
using a preallocated array. The difference in time is just a little over a nanosecond per 
sample on average, so use whichever leads to cleaner code for your use-case.

\begin{table}[t]
\caption{Sampling pairs of distinct integers: average time (ns)
with 99.9\% confidence intervals}\label{tab:pairs-ds}
\centering
\begin{tabular*}{\textwidth}{@{\extracolsep\fill}rrrr@{}}\toprule
$n$	& returning new array & preallocated array & Java record \\ \midrule
$n=16$   & $16.1 \pm 0.1$    & $13.7 \pm 0.2$ & $12.4 \pm {<}0.1$\\
$n=64$   & $15.5 \pm {<}0.1$ & $13.3 \pm 0.1$ & $12.0 \pm 0.3$ \\
$n=256$  & $15.6 \pm 0.5$    & $13.2 \pm 0.2$ & $11.8 \pm 0.1$ \\
$n=1024$ & $15.4 \pm 0.1$    & $13.2 \pm 0.4$ & $11.9 \pm 0.5$ \\
\bottomrule
\end{tabular*}
\end{table}

\subsection{Sampling Triples of Distinct Integers}

We now consider the results in Table~\ref{tab:triples} for sampling random triples of
distinct integers. As in the case of sampling pairs of distinct integers, we find that
the algorithm of this paper, \lstinline|RandomTriple| (Algorithm~\ref{alg:triple})
requires a constant time on average. The insertion sampling algorithm likewise has
an average time that is constant, although a higher constant than Algorithm~\ref{alg:triple}.
Both pool sampling and reservoir (R) sampling require time that increases linearly in
$n$, with reservoir sampling's times significantly higher than pool sampling due to 
generating significantly more random numbers during the sampling process for low $k$. 
Reservoir (L) sampling is also significantly slower than Algorithm~\ref{alg:triple}.

\begin{table}[t]
\caption{Sampling triples of distinct integers: average time (ns)
with 99.9\% confidence intervals}\label{tab:triples}
\centering
\begin{tabular*}{\textwidth}{@{\extracolsep\fill}rrrrrr@{}}\toprule
$n$	& \lstinline|RandomTriple| (Algorithm~\ref{alg:triple}) & insertion~\cite{cicirello2022applsci} & pool~\cite{Goodman1977} & reservoir (R)~\cite{Vitter1985} & reservoir (L)~\cite{Li1994} \\ \midrule
$n=16$   & $23.6 \pm {<}0.1$ & $41.8 \pm {<}0.1$ & $59.4 \pm 0.2$   & $179.3 \pm 0.3$   & $583.4 \pm 0.8$\\
$n=64$   & $22.4 \pm 0.3$    & $41.1 \pm 0.2$    & $91.3 \pm 0.3$   & $652.0 \pm 0.7$   & $1026.6 \pm 3.4$\\
$n=256$  & $22.4 \pm {<}0.1$ & $45.6 \pm 0.1$    & $286.0 \pm 1.1$  & $2113.4 \pm 23.6$ & $1459.3 \pm 103.8$\\
$n=1024$ & $21.8 \pm 0.1$    & $42.3 \pm 0.3$    & $1177.5 \pm 3.7$ & $8564.1 \pm 64.3$ & $1867.6 \pm 292.3$\\
\bottomrule
\end{tabular*}
\end{table}

In Table~\ref{tab:triples-ds} we further explore the effects of allocating a new array
for each sampled triple, versus using a preallocated array, versus returning a Java record.
The results in the triple of integers case follow the same pattern as that of the pairs of 
integers case.
For triples of distinct integers, returning a Java record is about 10\% faster
than using a preallocated array for the result, which in turn is about 10\% faster than
returning a new array.

\begin{table}[t]
\caption{Sampling triples of distinct integers: average time (ns)
with 99.9\% confidence intervals}\label{tab:triples-ds}
\centering
\begin{tabular*}{\textwidth}{@{\extracolsep\fill}rrrr@{}}\toprule
$n$	& returning new array & preallocated array & Java record \\ \midrule
$n=16$   & $23.8 \pm 0.1$    & $21.7 \pm {<}0.1$ & $19.2 \pm {<}0.1$ \\
$n=64$   & $23.2 \pm {<}0.1$ & $20.5 \pm 0.1$    & $18.0 \pm {<}0.1$ \\
$n=256$  & $23.4 \pm {<}0.1$ & $21.0 \pm {<}0.1$ & $18.7 \pm 0.2$ \\
$n=1024$ & $22.8 \pm 0.1$    & $20.0 \pm 0.5$    & $18.3 \pm 0.1$ \\
\bottomrule
\end{tabular*}
\end{table}

\section{Conclusions}\label{sec:conclusion}

In this article, we presented constant runtime algorithms for sampling pairs and
triples of distinct integers from the interval $[0,n)$. We conducted benchmarking
experiments with our Java implementations using JMH that demonstrate that these
sampling algorithms are substantially faster than using existing general purpose
algorithms for sampling $k$ integers from $[0,n)$. For example, reservoir (R)
sampling~\cite{Vitter1985} and pool sampling~\cite{Goodman1977} both require
linear time, with reservoir (R) sampling being especially computationally expensive
for low $k$ since it also requires $n-k$ random integers for sample size $k$.
Reservoir (L) sampling~\cite{Li1994} is superior to reservoir (R) for larger $n$,
but it is still very slow relative to the constant time algorithms
for the special cases of $k=2$ and $k=3$.
Although insertion sampling~\cite{cicirello2022applsci} has a constant runtime
for any fixed $k$ (e.g., runtime increases with $k$ but not with $n$), the 
special purpose algorithms for the two specific cases of $k=2$ and $k=3$ presented 
in this paper are significantly faster.

The structure of Algorithms~\ref{alg:pair} and~\ref{alg:triple} for sampling pairs and 
triples of distinct integers suggest a pattern for forming a special purpose algorithm
for any fixed $k$, such as the example provided for $k=4$ in Algorithm~\ref{alg:quad}.
Although it is unlikely practical or efficient for larger $k$ in software, it is 
possible that a circuit implementation of such a sampling network may have similar
benefits as sorting networks.

Our implementations of the algorithms for small random samples and three of the general
purpose sampling algorithms are available in the open source Java library $\rho\mu$. 
The experiments of this paper, including the fourth general purpose sampling algorithm, 
are also available on GitHub to assist with reproducibility.

\bmsection*{Conflict of interest}

The author declares no potential conflict of interests.

\bibliography{sampling}

\end{document}